\documentstyle[12pt]{article}

\newcommand{\vol}[1]{{\bf #1}}
\newcommand{\tselea}[1]{\label{#1} }
\newcommand{\tseleq}[1]{\label{#1} }
\newcommand{\tbib}[1]{\bibitem{#1}} %  {\bibitem{#1} [#1] }
\newcommand{\tref}[1] {(\ref{#1})} %{(\ref{#1}-#1)}
\newcommand{\tcite}[1]{\cite{#1}} % {\cite{#1} [#1] }
\newcommand{\tnote}[1]{}%{\footnote{#1}}
\newcommand{\tcaption}[2]{\vspace{#1} \caption{#2}}
\newcommand{\half}{\frac{1}{2}}
\newcommand{\bea}{\begin{eqnarray}}
\newcommand{\eea}{\end{eqnarray}}
\newcommand{\beann}{\begin{eqnarray*}}
\newcommand{\eeann}{\end{eqnarray*}}
\newcommand{\beq}{\begin{equation}}
\newcommand{\eeq}{\end{equation}}
\newcommand{\beqnn}{\begin{displaymath}}
\newcommand{\eeqnn}{\end{displaymath}}
\newcommand{\nnel}{\nonumber \\}
\newcommand{\npagepub} { }
\newcommand{\dash}[1]{{#1}\prime{}}
\newcommand{\alphabar}{\bar{\alpha}}
\newcommand{\gammabar}{\bar{\gamma}}
\newcommand{\taudash}{\dash{\tau}}
\newcommand{\set}[1]{\mbox{$\{ #1 \}$}}

\begin{document}
\typeout{--- Title page start ---}

\thispagestyle{empty}
\renewcommand{\thefootnote}{\fnsymbol{footnote}}

\typeout{--- preprint number and date ---}
\begin{tabbing}
\hskip 11.5 cm \= {Imperial/TP/91-92/36} %Preliminary version
\\
\> hep-ph/9209252 \\
\> 16th September, 1992 \\
\> (Printed out \today ) \\
%\hskip 1 cm \>{next preprint}\\
\end{tabbing}
\vskip 1cm
\begin{center}
{\Large\bf    A New Time Contour for Equilibrium Real-Time Thermal
Field Theories}
\vskip 1.2cm
{\large\bf T.S. Evans\footnote{E-mail: UMAPT85@UK.AC.IC.CC.VAXA}}\\
Blackett Laboratory, Imperial College, Prince Consort Road,\\
London SW7 2BZ  U.K.
\end{center}
\vskip 1cm
\begin{center}
{\large\bf Abstract}
\end{center}

A new time contour is used to derive real-time thermal field theory.
Unlike previous path integral approaches, no contributions to the
generating functional are dropped.

\vskip 1cm
\begin{center}
\end{center}

\vskip 1cm
\npagepub
\typeout{--- Main Text Start ---}

%See pp.175
\renewcommand{\thefootnote}{\arabic{footnote}}
\setcounter{footnote}{0}

In the path integral approach to equilibrium thermal
field theories, one is calculating thermal expectation values,
typically
\begin{eqnarray}
\lefteqn{\Gamma_C (\tau_1,\tau_2,...,\tau_N) =} \nnel && Tr\{e{-
\beta H}
T_C \phi_1(\tau_1)\phi _2(\tau_2)...\phi_N(\tau_N) \}
 / Tr\{e{- \beta H}\}  .
\tselea{eftgf}
\end{eqnarray}
Here $T_c$ indicates that the fields are path-ordered with respect to
the order along a path, $C$,
of their complex time arguments \set{\tau} \tcite{NS,LvW}
\tnote{LvW, 2.1.13; TSEnpt, 2.2}
with appropriate sign changes when fermionic fields are involved
\tnote{TSEnpt,5.1,2.6}.   Throughout we shall suppress spinor and
other
indices as well as dependence on spatial variables where necessary.
The
results are not altered by their inclusion.  The $e{-\beta H}$ factor
is used to shift the bra states in the trace by $- \imath \beta$
in time so that all times involved lie on a path, $C$, which must run
in the complex
time plane so that it ends $-\imath \beta$ below its starting point.

Another limitation on $C$ that is often imposed comes from
studying the behaviour of the Green functions.
Inserting complete sets of energy eigenstates into the
thermal Wightman function, say $\Gamma_{a_1...a_N}$, where the
$N$ fields are in a fixed order, we have
\begin{eqnarray}
\lefteqn{\Gamma_{a_1...a_N}(\tau_1,...,\tau_N) } \nnel
&=&
Tr\{e{- \beta H}
\phi _{a_1}(\tau_{a_1})\phi_{a_2}(\tau_{a_2})...\phi_{a_N}(\tau_{a_N})
\}
 / Tr\{e{- \beta H}\}.
\tselea{etwfdef}
\nnel &=& (-1)p
\sum_{m_1,m_2,...,m_N} [  exp\{-\imath E_{m_1}(b
-\tau_{a_1}+\tau_{a_N})\} .  \nnel
&&  \left( \prodN_{j=2} exp\{
- \imath E_{m_j}(-\tau_{a_j}+\tau_{a_{j-1}})\} \right) . \left(
\prodN_{j=1} \langle m_j|\phi_{a_j}(0)|m_{j+1} \rangle
 \right) ]  / Tr\{e{- \beta H}\} .
\nonumber \\
\tselea{etwfees}
\end{eqnarray}
The
$\{|m_j \rangle\}$ are a complete set of energy eigenstates for
temperature $\beta{-1}$ ($m_{1+N}=m_1$).
We assume that the energy sums in \tref{etwfees} are uniformly
convergent when the exponentials in \tref{etwfees} are of the form
$exp\{z_jE_j\}$ with $\Re e\{z_j\}<0$.  The
thermal Wightman functions, $\Gamma_{a_1a_2....a_3}(\{\tau\})$, are
therefore bounded when the complex time arguments $\tau_j$
satisfy $\{\tau\} \in A_{a_1a_2....a_3}$ \tcite{TSEnpt} where
\begin{eqnarray}
A_{a_1a_2...a_N} &:=&
\{\{\tau_{a_1},\tau_{a_2},...,\tau_{a_N}\} \, |  Im(\tau_{a_N})\geq
Im(\tau_{a_{N-1}}) \geq ...  \nnel
&& \mbox{      }...\geq Im(\tau_{a_1})\geq Im(\tau_{a_N}- \imath
\beta) \} .
%\tselea{eadef}
\end{eqnarray}
This then implies that the thermal
Wightman function is also analytic in this region \tcite{Hor}.

The path-ordered Green functions \tref{eftgf}
are merely a combination of products of
these thermal Wightman functions \tref{etwfdef}
with path ordering theta functions. %\tcite{LvW,NS,KK}.
In order to ensure
that the path ordered Green function $\Gamma_c(\{\tau\})$ is also
analytic away from hypersurfaces of equal time arguments, one finds
that
one must
ensure that the time curve $C$ always has a decreasing imaginary part
as
one moves along $C$ \tcite{TSEnpt}.
This is a further condition that one imposes on the
curve $C$ (though a constant real part may be acceptable if one is
prepared to accept generalised rather than analytic functions,
see section 2.1 in \tcite{LvW}\tnote{LvW, above 2.1.21}).

Given these two limitations on $C$ one is then free to choose the time
curve to suit the problem at hand.  The standard choice is to choose
one
running parallel to the imaginary axis, $C=C_I$ of fig.\tref{fc}.
This
gives the ITF (Imaginary Time-Formalism) or Matsubara method.  Because
this has no real times present, one must make an analytic continuation
to real times of the Green functions calculated if one wishes to
obtain
dynamical information.  This operation is in principal highly
non-trivial because the analytic structure at finite temperature is
much
more complicated than at zero temperature.

To avoid this problem RTF (Real-Time Formalisms) were developed.  The
path integral approach was highlighted by Niemi and Semenoff
\tcite{NS}
though RTF had also been considered earlier (see \tcite{LvW} for a
review).  In these approaches one chooses
the curve in the complex time plane to be
$C=C_R$ where $C_R=C_1 \oplus C_2 \oplus C_3 \oplus
C_4$ runs from $-T+\tau_0$ to $+T+\tau_0$ ($C_1$),
$+T+\tau_0$ to $+T+\tau_0-\imath \bar{\alpha} \beta$ ($C_2$),
$+T+\tau_0-\imath \bar{\alpha} \beta$ to
$-T+\tau_0-\imath \bar{\alpha} \beta$ ($C_3$), and
$-T+\tau_0-\imath \bar{\alpha} \beta$ to
$-T+\tau_0-\imath \beta$ ($C_4$) \tcite{NS,LvW,Raybook}.
This is shown in Figure \ref{fc}.
The parameters $\alpha=1-\bar{\alpha}$ and $\tau_0$ is
arbitrary and reflect some of the freedom of
choice in the path $C$.  Physical results do not therefore
depend on them.  The limit of $T\rightarrow \infty$ is taken to
produce
the standard RTF.

The draw back of this approach is that one has to show that
the vertical sections $C_3$,$C_4$ are irrelevant
\tcite{NS,LvW,Raybook}.
Otherwise one would have an unwieldy combination with few similarities
to zero temperature Minkowskii space field theory, and one would
not see the close relationship between this path integral method and
the Thermo Field Dynamics and $C\ast$-algebra approaches to RTF (see
\tcite{LvW} for a review).
To see that $C_3$,$C_4$ can be ignored, one has to
take the $T \rightarrow \infty$
limit and then demand that the sources, $j(\tau)$, are zero at
$\Re e\{ \tau \} \rightarrow \infty$ in manner reminiscent of the
usual
asymptotic conditions used in zero temperature field theory.

Dropping these contributions is disquieting to many.  The use of an
asymptotic condition at infinite times in the past and future must be
some sort of approximation in thermal field theories.
Any real-time Green function is telling us
about dynamics and therefore some sort of deviation from equilibrium.
To use equilibrium theories in such problems, we are assuming that
deviation from equilibrium is `small' and can be ignored on
`short' time scales.  After a finite time one must face the fact
that the system is
returning to equilibrium.  This can not be done within equilibrium
thermal theories.  Thus one should not really be relying on what the
theory is doing on infinite time scales.

In practice, calculations of real-time Green functions in RTF and
ITF seem to be in complete agreement when we compare the same type of
Green functions (the same diagram usually corresponds to different
Green
functions in different formalisms \tcite{TSEnpt,TSEo,Ko}).
The one area that still appears to be a problem is when one is
interested in static (zero energy) Green functions such as appear in
free energy calculations.  In RTF, some of the relevant diagrams are
infinite due to the appearance of singularities of the form
$[\delta(k2-m2)]{N\geq2}$ \tcite{TSEwin} ({\it pace}
\tcite{LvW,BD}).  Extra rules were shown to be needed for RTF
calculations in these situations \tcite{MNU,TSEz}.  In the path
integral derivation of these rules \tcite{TSEz} the $C_3$,$C_4$
sections did contribute when static sources were involved.
Only by using a simple
trick leading to an extra Feynman rule could one use the standard RTF
method.

It would therefore be advantageous to find a path integral approach to
RTF that did not involve dropping any parts yet gave the standard RTF
Feynman rules.  The starting point is to note that it is common to
think of the $C_1$,$C_2$ of the standard RTF curve $C_R$ of
fig.\tref{fc} as having infinitesimal slopes downwards so that the
path ordered Green functions defined on these curves are analytic.
The difference in the imaginary part of the ends of the $C_1$,$C_2$ curves
is also infinitesimal.  This latter property is unnecessary for the Green
functions to be well behaved and it is this condition which
we drop.  Thus we might
try the new curve $C_N = C_{N1} \oplus C_{N2}$ shown in fig.\tref{fnc}.
This curve satisfies all the requirements discussed above.
It is convenient to parameterise the curve using a real variable, $t$,
and a thermal label, $a=1,2$.
We define a complex valued doublet, $\taua$, so that
\begin{equation}
\taua(t)= \protect\left\{
\begin{array}{lcl}
g1 t+\tau_0 &\in C_{N1} & \mbox{ if $a=1$}
\\
g2 t-\imath \alphabar \beta + \tau_0 & \in C_{N2} & \mbox{ if $a=2$}
\end{array}
\right.
\tseleq{enrtfparam}
\end{equation}
The constants $g1,g2$ are
\beq
g1= 1 - \imath \frac{\gamma \beta}{\alpha T}; \; \;
g2= -1 - \imath \frac{(\gamma - \alphabar) \beta}{\alpha T}.
\tseleq{egdef}
\eeq
Thus $C_N$ runs from $\tau1(-\alpha T) = -\alpha T+ \imath \gamma
\beta
+ \tau_0$ to
$\tau1(\alphabar T)=
\alphabar T + \imath \gamma \alphabar \beta / \alpha$
($C_{N1}$) and then from
$\tau2(\alphabar T) =
\alphabar T + \imath \gamma \alphabar \beta / \alpha$
to $\tau2(-\alpha T)=-\alpha T+ \imath \gamma \beta + \tau_0$
($C_{N2}$).
The $\tau_0$ is an arbitrary
complex time and it is in the region of $\tau_0$ that we wish to
investigate the Green functions.  The ends of the curve must be
separated by $-\imath \beta$ so that we have $\gamma + \gammabar =1$,
though the Feynman rules will turn out to be completely independent of
this $\gamma$ parameter.
We have also defined $\alpha + \alphabar =1$ which will turn out to
play
the same role for $C_N$ as they did for $C_R$.

The constant $T$ is a
time and one takes the limit $T\rightarrow \infty$
so that the slope of the two sections
$C_{N1}$,$C_{N2}$ becomes infinitesimal as required.  Also in this
limit, $C_N$ and $C_R$ are then
essentially identical in the region $|\tau - \tau_0| \ll T$.
The limit of $T\rightarrow \infty$ will
therefore generate the usual RTF method and yet no part of the contour
has been dropped.  There can be no suspicion that RTF contains
anything
less than any other formalism such as ITF where a different curve is
employed.

It is enlightening to check this new approach to RTF by studying the
simple example of a real relativistic self-interacting field theory
with
Lagrangian
\beqnn
{\cal L} = \half (\partial\mu \phi)2 + \half m2 \phi2  - V[\phi].
\eeqnn
The generating functional can be written as
\beann
Z[j] &=& exp\{V[-\imath \frac{\partial}{\partial j} ] \} \; . \;
Z_0[j]
\nnel
Z_0[j] &=& exp \{ - \frac{\imath}{2} \int_C d\tau d\taudash
j(\tau) \Delta_c(\tau - \taudash) j(\taudash) \}
\eeann
where
\beq
-( \Box_c + m2)\Delta_c(\tau - \taudash) = \delta_c(\tau - \taudash).
\tseleq{ekg}
\eeq
The $\Box_c$ contains time derivatives taken along $C$.  Also
$\theta_c(\tau - \taudash)=1$ if $\tau$ is further along $C$
than $\taudash$ otherwise $\theta_c(\tau - \taudash)=0$,
$\frac{\partial}{\partial \tau} \theta_c(\tau - \taudash) =
\delta_c(\tau - \taudash)$ and $\int_C \delta_c(\tau - \taudash) = 1$
defines the delta function (c.f. \tcite{NS,LvW,KK}).

Using the KMS condition and the equal time commutation relation
$[\phi(\tau,\vec{x}),\phi(\tau,\vec{y})] = 0 $ as
boundary conditions the solution of \tref{ekg} is
\beq
\Delta_c(\tau,\omega) = \frac{- \imath}{2 \omega}
\frac{1}{1-e{- \beta \omega} } \{
[e{-\imath \omega \tau} + e{-\beta  \omega } e{+\imath \omega
\tau}] \theta_c(\tau) +
[e{+\imath \omega \tau} + e{-\beta  \omega } e{-\imath \omega
\tau}] \theta_c(-\tau)          \}
%\tseleq{eprop}
\eeq
where $\omega= |(\vec{k}2 +m2)\half|$ and we have taken the Fourier
transform with respect to the spatial variables.

Now consider $C=C_N$ and define doublet fields and sources with real
time arguments $\{t\}$ in terms of the original fields and sources, whose
time arguments lie on $C_N$, through
\beq
\phia(t) = \phi(\taua(t)) ; \; \; ja(t) = ga \; j(\taua(t)).
\tseleq{edoubdef}
\eeq
The $a=1,2$ is the thermal label, and the $\taua$ is the
parameterisation of the new curve given in \tref{enrtfparam}.
The generating functional can then be written in the usual final RTF
form
\beann
Z[j] &=& exp\{ \sum_{a=1,2} \int_{-\alpha T}{\alphabar T} dt \;
ga \; V[-\imath \frac{\partial}{\partial ja} ] \}
.Z_0[j]
\nnel
Z_0[j] &=& exp \{ - \frac{\imath}{2} \sum_{a,b=1,2}
\int_{-\alpha T}{\alphabar T} dt \; d\dash{t} \;
ja(t) \Delta{ab}(t-\dash{t}) jb(t) \}
\eeann
but where in this case no term has been dropped or absorbed into the
normalisation.  The two-by-two matrix propagator is defined to be
\beqnn
\Delta{ab}(t-\dash{t}) = \Delta_c(\taua(t)-\taub(\dash{t}))
\eeqnn
The Feynman rules using these definitions lead to a factor of $ga$
(as defined in \tref{egdef})
associated with each vertex.  In the $T \rightarrow \infty$ limit this
leads to the usual additional factor of $-1$ appearing in type two
vertices
as compared with the type one vertices.
The definition of the doublet fields and sources in \tref{edoubdef}
is fairly arbitrary and
other useful definitions can be used which lead to variations in the
Feynman rules \tcite{KK}.

When the limit $T \rightarrow \infty$ is taken the doublet fields are
then
\beann
\lim_{T \rightarrow \infty} \phi1(t) &=& \phi(t) \nnel
\lim_{T \rightarrow \infty} \phi2(t) &=& \phi(t-\imath\alphabar\beta)
\eeann
where without loss of generalisation we have chosen $\tau_0$=0.  Thus
we are indeed calculating the same Green functions as in standard RTF,
for example
\beqnn
\frac{\partialN Z[j] }{\partial j1(t_1) ... j1(t_N)} =
Tr\{e{- \beta H}
T \phi_1(\tau_1)\phi _2(\tau_2)...\phi_N(\tau_N) \}
 / Tr\{e{- \beta H}\}
\eeqnn
which is the connected time-ordered Green function
of physical real-time fields.

Calculating the Fourier transform of the propagator in this limit
requires care with the various
infinitesimal quantities.  This is because
\bea
\lefteqn{\lim_{T \rightarrow \infty}
\int_{-T}{T} dt \; e{\imath (p + \imath \epsilon) t}
\; . \;
\theta(t)e{-\imath \omega . (1-\imath \eta )t} }
\nnel
&=& \frac{ \imath}{p - \omega + \imath \epsilon + \imath \omega \eta }
[ 1 - \lim_{T \rightarrow \infty}
(e{ \imath (p-\omega)T} e{ -(\epsilon + \omega
\eta) T }) ]     .
\tselea{eft}
\eea
where $p$ and $\omega$ are real and $\epsilon$ and $\eta$ are real and
infinitesimal.
Consider the calculation of $\Delta{11}$.
Then $\eta=(1-g1)/\imath$ is positive
and comes from the infinitesimal gradient given
to the otherwise horizontal $C_{N1}$.  As usual if only particle
solutions are propagating at large times, this infinitesimal
slope to the time path plays the same role as the usual
$\imath \epsilon$ shift in the energy used to regulate the integral.
However at non-zero temperatures we have both particle and anti
particle
solutions, and this means we also have terms where $\omega \rightarrow
-\omega$ in \tref{eft}.  The slope of the time curve is of course the
same so now we require that $\epsilon > \eta $ for a finite result.
The same behaviour is found when looking at the particle contributions
to the
$\theta(-t)$ part of the propagator and so on
for all parts of the matrix RTF propagator.  In general
we find
that the infinitesimal slope given to time curves at non-zero
temperature
is {\em not} sufficient to guarantee convergence of the Fourier
transform of the free propagator when $T \rightarrow \infty$.

The problem is easily avoided. We are
still free to give the time curves infinitesimal slopes but we must
also have the usual $\imath \epsilon$ shift in the energies.  We must
then take the slopes to zero before we take the $\epsilon\rightarrow
0$
limit.  In terms of the RTF curves (both for this new approach and for
the old style curves) we must take the $T\rightarrow \infty$ limit
before the $\epsilon \rightarrow 0$ limit.

With this proviso we find that the usual RTF propagators in momentum
space are reproduced.
In this example where the definitions have been chosen to give a
factor
of $-1$ appearing in the type two vertices over the type one, we find
that
\beann
\imath \Delta{11}(k_0,\omega) &=&
\frac{\imath}{k_02-\omega2+\imath \epsilon}
+ \frac{1}{e{ \beta \omega} -1} 2\pi \delta(k_02-\omega2)
\\
\imath \Delta{12}(k_0,\omega) &=&
\frac{e{ \beta \omega/2}}{e{ \beta \omega} -1}
e{ (\alphabar-\half)\beta k_0 /2}
\; 2\pi \delta(k_02-\omega2)
\\
\imath \Delta{21}(k_0,\omega) &=&
\frac{e{ \beta \omega/2}}{e{ \beta \omega} -1}
e{ - (\alphabar-\half)\beta k_0 /2}
\; 2\pi \delta(k_02-\omega2)
\\
\imath \Delta{22}(k_0,\omega) &=&
\frac{-\imath}{k_02-\omega2 - \imath \epsilon}
+  \frac{1}{e{ \beta \omega} -1} 2\pi \delta(k_02-\omega2)
\eeann

The physical origin of these problems is that $T$ represents the time
scale over which the system starts to come into equilibrium.  In
looking
at dynamical information in the form of real-time Green functions, we
have the inherent contradiction that we must be pushing the system
out of equilibrium yet we assume the system is in equilibrium for
all time.  This is a good approximation on time scales less than $T$
but is bad otherwise.  Mathematical consistency therefore demands that
this
characteristic time scale is taken to infinity in such equilibrium
schemes.  It is not surprising that the $T\rightarrow \infty$ limit
must
be taken first.

This also explains why we should not be surprised if when using RTF to
look at
zero energy Green functions we have to work a little harder as these
involve
averages over all times including those of order $T$.  This could be
seen in the extra rule found to be necessary \tcite{MNU,TSEz}.
When using the
standard RTF time curve the vertical sections were shown to be giving
all the contribution to the generating functional \tcite{TSEz} but
after some manipulations, it was shown the standard RTF approach could
be used if one extra Feynman rule was added.

For the new time curve the
formal manipulations of \tcite{TSEz} work exactly as before.
The ends of the curves are
significant and so the infinitesimal slope can not be ignored.
The same trick as before allows
one to take account of this when using the new curve by introducing
the
same extra rule.  Once again the new time curve completely reproduces
the
standard RTF method without dropping any contributions.  This gives us
further confidence that ITF and RTF contain the same physical
information.

I would like to thank R. Kobes, G. Kunstatter and the Institute for
Theoretical Physics at the University of Winnipeg for their
hospitality and the Central Research Fund of the
University of London for financial support for my visit to
Winnipeg.

\npagepub

\typeout{NRTF - references}

\npagepub

% *** Figure fc ***
\typeout{figure: Standard RTF and ITF curves}
\begin{figure}[thb]

\begin{center}
\setlength{\unitlength}{1in}
%use 1 inch for paper, 0.8in for preprint
\begin{picture}(5,5.5)

%Finished on Jan. 16, 1991
%  This Picture is taken from the paper
%  The Heisenberg and interaction representations in thermo
%  field dynamics.

%units are inches

\thicklines
%  *** axes ***
\put(-0.5,4.6){\vector(1,0){6.0}}
\put(5.0,4.7){\large $\Re e \, (\tau-\tau_0) $ }
\put(2.5,0.0){\vector(0,1){5.0}}
\put(2.6,5.0){\large $\Im m \, (\tau-\tau_0) $ }

\thinlines

% *** RTF curve ***

\put(0,4.5){\vector(1,0){3.0}}
\put(3.0,4.5){\line(1,0){2.0}}

\put(5,4.5){\vector(0,-1){0.5}}
\put(5,4.0){\line(0,-1){0.5}}

\put(5,3.5){\vector(-1,0){2.0}}
\put(3.0,3.5){\line(-1,0){3.0}}

\put(0,3.5){\vector(0,-1){1.5}}
\put(0,2.0){\line(0,-1){1.5}}

\put(2.8,3.7){\makebox(0,0){\large $-\imath \alphabar \beta$}}
\put(2.8,0.5){\makebox(0,0){\large $-\imath \beta$}}
\put(3.0,4.2){\large $C_1$}
\put(3.0,3.2){\large $C_2$}
\put(5.1,4.1){\large $C_3$}
\put(0.2,2.0){\large $C_4$}

%  *** ITF curve ***

\put(2.3,4.5){\vector(0,-1){2.0}}
\put(2.3,2.5){\line(0,-1){2.0}}
\put(1.8,1.8){\large $C_I$}

\end{picture}

\end{center}

\tcaption{3cm}{The time paths used for RTF and ITF.}

\label{fc}

\end{figure}

% *** Figure fnc ***
\typeout{Figure: New RTF curve}
\begin{figure}[thb]

\begin{center}
\setlength{\unitlength}{0.8in}
\begin{picture}(6,9)(-2,-8) % move origin

\thicklines
%  *** axes ***
\put(-2.0,0){\vector(1,0){5.0}}
\put(2.5,0.15){\large $\Re e \, (\tau - \tau_0 )$}
\put(0,-7.0){\vector(0,1){9.0}}
\put(0.1,1.7){\large $\Im m \, (\tau - \tau_0)$}

\thinlines

% *** New RTF curve ***

\put(-1.0,1.0){\vector(1,-1){3.0}}
\put(2,-2){\line(1,-1){1.0}}
\put(2,-1.9){\large $C_{N1}$}

\put(3,-3){\vector(-1,-1){1.0}}
\put(2,-4){\line(-1,-1){3.0}}
\put(2,-4.3){\large $C_{N2}$}

\put(0.1,0.1){\large $0$}
\put(0.1,-6.1){{\large $-\imath \alphabar \beta$}}

\put(-2.0,1.1){\large $-\alpha T + \imath \gamma \beta$}
\put(-1.5,-6.7){\makebox(0,0){\large $-\alpha T -\imath \gammabar
\beta$}}
\put(3.1,-3.0){{\large $\alphabar T -\imath \gamma \alphabar \beta /
\alpha$}}

\end{picture}

\end{center}

\tcaption{1cm}{The new time path for RTF.}

\label{fnc}

\end{figure}

\end{document}